\documentclass{article}

\usepackage{graphicx}

\begin{document}

\title{{\bf Spectropolarimetric Observations of Active Galactic Nuclei with the 6-m BTA Telescope.}
\\ {\small \it (Published in Astronomy Letters, 2011, Vol. 37, No. 5, pp. 302-310)}}

\author{V.L. Afanasiev$^1$, N.V. Borisov$^1$, Yu.N. Gnedin$^2$\thanks{E-mail: gnedin@gao.spb.ru},
\\ T.M. Natsvlishvili$^2$, M.Yu. Piotrovich$^2$ and S.D. Buliga$^2$.
\\ (1) Special Astrophysical Observatory, Nizhnii Arkhyz, Russia.
\\(2) Central Astronomical Observatory at Pulkovo, St.-Petersburg, Russia.}

\maketitle

\begin{abstract}

We present the results of our spectropolarimetric observations for a number of active galactic nuclei (AGNs) carried out at the 6-m telescope with the SCORPIO focal reducer. The derived wavelength dependences of the polarization have been analyzed by taking into account the Faraday rotation of the polarization plane on the photon mean free path in a magnetized accretion disk. As a result, based on traditional accretion disk models, we have determined the magnetic field strength and distribution and a number of physical parameters of the accreting plasma in the region where the optical radiation is generated.

\end{abstract}

\section{Introduction}

Investigation of active galactic nuclei (AGNs) and quasars is one of the central problems in modern astronomy. According to present views, a central supermassive black hole and an accretion disk around the central engine are the main objects of AGNs. The accretion disk has a complex structure and quite a few theoretical works, starting from the classical work by Shakura and Sunyaev (1973), are devoted to the description of its structure.

Naturally, a decisive contribution to the solution of this problem belongs to observations. Such observations are carried out at many astronomical observatories and with space telescopes. As a rule, the latter observe the X-ray emission from AGNs and quasars.

Among the main methods of observing these objects, a prominent role belongs to spectropolarimetry. At the Special Astrophysical Observatory, the SCORPIO focal reducer (Afanasiev and Moiseev 2005), which can also operate in the spectropolarimetric mode, is used for this purpose at the 6-m BTA telescope. This instrument can measure the linear and circular polarization of starlike objects in a wide spectral range with an accuracy of 0.2-0.3\%. The total quantum efficiency of such observations can reach 30\%.

The physical conditions in accretion disks around supermassive black holes essentially do not allow a direct method of magnetic field measurement, the method of Zeeman spectropolarimetry, to be applied. Therefore, when analyzing our spectropolarimetric observations, we use an indirect method of magnetic field determination developed by Gnedin and Silant'ev (1997), Gnedin et al. (2006), and Dolginov et al. (1995). The idea of the method is that if the Faraday rotation on the photon mean free path in the process of scattering by electrons is taken into account, then the degree of polarization and the position angle, along with their dependences on the wavelength of the accretion disk emission, are completely determined by the distribution, i.e., geometry, of the magnetic field within the accretion disk. The polarization turns out to be lower than that obtained by Sobolev (1949) and Chandrasekhar (1950) by solving the problem of multiple light scattering in a plane-parallel atmosphere. This difference is related to the Faraday depolarization of the emission as it is scattered in the accretion disk. In recent years, polarimetric observations have increased greatly in importance, because they allow a decisive choice to be made among an increasing number of accretion disk models. This increase is attributable to an increase in the number of numerical simulations of the accretion disk structure. By applying this technique, we will determine the magnetic field strengths in the accretion disk region where the observed AGN radiation is generated, along with the exponent of the power-law magnetic field distribution in the disk itself.

\section{Results of our observations at the 6-m BTA telescope in the SCORPIO spectropolarimetric mode}

We performed the spectropolarimetric observations of a sample of AGNs at the 6-m BTA telescope of the Special Astrophysical Observatory in 2008. 2009 during three sets of observations. For our observations, we selected objects with published mass estimates for their central black holes $M_{BH}$. The observations were carried out with the SCORPIO focal reducer in the spectropolarimetric mode mounted at the prime focus. We used an EEV42-40 2048$\times$2048 pixel CCD array with a pixel size of 13.5$\times$13.5 $\mu$m as the detector and a VPHG550g volume holographic phase grating from the SCORPIO kit operating in the range 3500.7200 \AA as the dispersing element. The reciprocal linear dispersion in the detector plane was 1.8 \AA /pixel. In the spectrograph, we used a set of five circular diaphragms 4''.5 in diameter arranged in the form of a pseudoslit with a step of 9.7-arcsec. A Savart plate placed behind the diaphragms was used as the polarization analyzer. We used the central diaphragm to take the spectra of an object in perpendicular polarization planes and the remaining diaphragms to take the night-sky spectra. The actual spectral resolution of our data was determined by the monochromatic image of the diaphragms and was 40.42 \AA. The seeing in all sets of observations was at least 2''.

The technique of polarization observations and calculations was described by Afanasiev et al. (2005). To calibrate the wavelengths and the relative transmission of the diaphragms, we used an Ar-Ne-He filled line-spectrum lamp and a quartz lamp. To calibrate the spectropolarimetric channel of the spectrograph, we observed standards from Turnshek et al. (1990). The data were processed and analyzed by a standard technique using specialized software packages written in the IDL6.2 environment. Figure 1 shows an example of a typical presentation of the processing results for one of the observed objects.

The list of observed objects and the main results of our observations are presented in Table 1. It gives the object names, AGN magnitudes in the $V$ band ($m_V$), redshifts $z$, types of AGN, dates of observations, exposure times $T_{exp}$, and mean values of the linear polarization $P_V$ and position angle of the polarization plane $PA_V$ in the $V$ band. The last column in the table gives the exponent (index) $n$ in the wavelength dependence of the linear polarization $P_l(\lambda) \sim \lambda^n$. The error in the position angle did not exceed $2-3^{\circ}$. When calculating n by the least-squares method, we rejected the polarization values in the region of emission lines. The values of n obtained will be used below to determine the power law of the radial magnetic field distribution in the accretion disk.

\begin{figure}
 \centerline{\includegraphics[height=22cm]{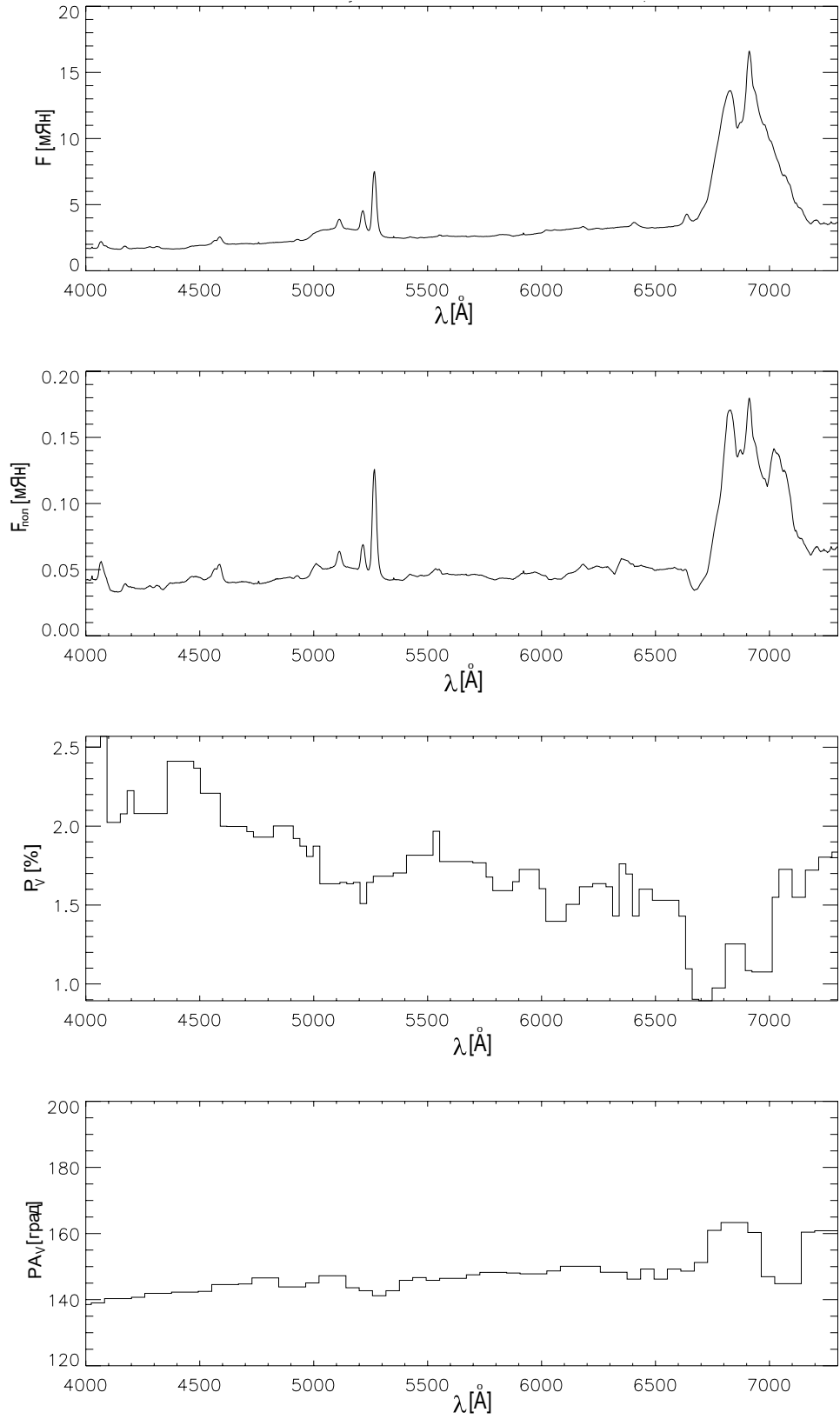}}
 \caption{Polarization measurements of 3C 390.3.}
\end{figure}

\section{Correlation between polarization parameters and characteristics of the central black hole}

It is interesting to compare our values of the polarization in continuum and the exponent in the wavelength dependence of the polarization $P_l(\lambda) \sim \lambda^n$ with black hole parameters. Table 2 lists the black hole luminosities and masses obtained by other authors from optical and ultraviolet spectroscopic observations (see the notes to Table 2). Analysis of Table 2 shows that there is a weak correlation between the polarization and black hole mass and a noticeable correlation between the exponent in the wavelength dependence of the polarization and black hole mass (Fig.2).

\begin{table}
 \footnotesize
 \caption[]{Results of our observations.}
 \centering
 \begin{tabular}{l c c c c c c c c}
 \hline
 Object      &$m_v$ & z     & Type  & Date     &$T_{exp}$, s & $P_V[\%]$    & $PA_V [deg]$& n \\
 \hline
 PG 0007+106 & 15.2 & 0.089 & Sy1   & 30.11.08 & 3000 & 1.02$\pm$0.38 & 83  &  0.15$\pm$0.25 \\
 PG 0026+129 & 15.3 & 0.142 & QSO   & 30.11.08 & 3000 & 1.07$\pm$0.28 & 99  & -0.45$\pm$0.33 \\
 PG 0049+171 & 16.1 & 0.064 & Sy1.5 & 24.09.09 & 2160 & 1.42$\pm$0.31 & 247 & -0.28$\pm$0.18 \\
 PG 0157+001 & 15.7 & 0.163 & Sy1.5 & 01.12.08 & 3000 & 0.78$\pm$0.28 & 17  & -0.52$\pm$0.28 \\
 PG 0804+761 & 14.7 & 0.100 & QSO   & 02.12.08 & 3000 & 1.00$\pm$0.38 & 83  &  0.24$\pm$0.38 \\
 PG 0844+349 & 14.5 & 0.064 & Sy1   & 29.11.08 & 3000 & 0.85$\pm$0.10 & 243 & -1.17$\pm$0.17 \\
 PG 0953+414 & 15.3 & 0.234 & QSO   & 03.12.08 & 3000 & 0.39$\pm$0.12 & 317 &  0.11$\pm$0.13 \\
 PG 1022+519 & 15.8 & 0.045 & Sy1   & 30.11.08 & 3000 & 0.83$\pm$0.30 & 259 & -2.37$\pm$0.45 \\
 PG 1116+215 & 14.4 & 0.177 & QSO   & 29.11.08 & 3000 & 0.57$\pm$0.12 & 193 & -1.26$\pm$0.13 \\
 PG 2112+059 & 15.9 & 0.466 & QSO   & 29.11.08 & 3000 & 1.04$\pm$0.21 & 258 &  0.45$\pm$0.17 \\
             &      &       &       & 18.08.09 & 3600 & 1.08$\pm$0.20 & 243 &  0.35$\pm$0.10 \\
 PG 2130+099 & 14.7 & 0.063 & Sy1   & 30.11.08 & 3000 & 0.62$\pm$0.15 & 53  & -0.05$\pm$0.32 \\
 PG 2209+184 & 15.9 & 0.070 & Sy1   & 24.09.08 & 3600 & 0.83$\pm$0.29 & 200 & -0.75$\pm$0.21 \\
 PG 2214+139 & 15.1 & 0.066 & Sy1   & 28.11.08 & 3000 & 1.58$\pm$0.18 & 323 & -0.69$\pm$0.15 \\
 PG 2233+134 & 16.3 & 0.326 & QSO   & 29.11.08 & 3000 & 0.67$\pm$0.23 & 253 &  0.28$\pm$0.28 \\
 3C 390.3    & 15.2 & 0.056 & Sy1   & 29.11.08 & 3000 & 2.09$\pm$0.22 & 140 & -0.57$\pm$0.22 \\
             &      &       &       & 17.08.09 & 3600 & 1.58$\pm$0.18 & 146 & -0.64$\pm$0.07 \\
             &      &       &       & 24.09.09 & 3600 & 1.80$\pm$0.24 & 144 & -0.58$\pm$0.06 \\
 \hline
 \end{tabular}
\end{table}

\begin{table}
 \footnotesize
 \caption[]{Masses of the central black holes and polarization in continuum}
 \centering
 \begin{tabular}{l l c c c c c c}
 \hline
 Object & Type& $\log{\lambda} L_{\lambda}$ & $\log{\frac{M_{BH}}{M_{\odot}}}$ & Ref.   & $P_V$[\%] & $n$ & Ref. \\
        &     & [erg/s] (opt.)              &                                  &        &           &     &      \\
 \hline
 PG 0007+106 & Sy1   & 44.82 & $8.73_{-0.10}^{+0.08}$ & 6 & 1.02$\pm$0.38 &  0.15$\pm$0.25 & 1 \\
 PG 0026+129 & QSO   & 45.02 & $8.59_{-0.12}^{+0.07}$ & 7 & 1.07$\pm$0.28 & -0.45$\pm$0.33 & 1 \\
 PG 0049+171 & Sy1.5 & 44.00 & $8.35_{-0.10}^{+0.08}$ & 6 & 1.42$\pm$0.31 & -0.28$\pm$0.18 & 1 \\
 PG 0157+001 & Sy1.5 & 44.98 & $8.17_{-0.10}^{+0.08}$ & 6 & 0.78$\pm$0.28 & -0.52$\pm$0.28 & 1 \\
 PG 0804+761 & QSO   & 44.94 & $8.84_{-0.06}^{+0.05}$ & 7 & 1.00$\pm$0.38 &  0.24$\pm$0.38 & 1 \\
 PG 0844+349 & Sy1   & 44.35 & $7.97_{-0.23}^{+0.15}$ & 7 & 0.85$\pm$0.10 & -1.17$\pm$0.17 & 1 \\
 PG 0953+414 & QSO   & 45.40 & $8.42_{-0.10}^{+0.08}$ & 6 & 0.39$\pm$0.12 &  0.11$\pm$0.13 & 1 \\
 PG 1022+519 & Sy1   & 43.70 & $7.15_{-0.11}^{+0.09}$ & 6 & 0.83$\pm$0.30 & -2.37$\pm$0.45 & 1 \\
 PG 1116+215 & QSO   & 45.40 & $8.53_{-0.10}^{+0.08}$ & 6 & 0.57$\pm$0.12 & -1.26$\pm$0.13 & 1 \\
 PG 2112+059 & QSO   & 46.18 & $9.00_{-0.11}^{+0.09}$ & 6 & 1.06$\pm$0.21 &  0.40$\pm$0.15 & 1 \\
 PG 2130+099 & Sy1   & 44.46 & $8.66_{-0.06}^{+0.05}$ & 7 & 0.62$\pm$0.15 & -0.05$\pm$0.32 & 1 \\
 PG 2209+184 & Sy1   & 44.47 & $8.77_{-0.10}^{+0.08}$ & 6 & 0.83$\pm$0.29 & -0.75$\pm$0.21 & 1 \\
 PG 2214+139 & Sy1   & 44.66 & $8.55_{-0.12}^{+0.09}$ & 6 & 1.58$\pm$0.18 & -0.69$\pm$0.15 & 1 \\
 PG 2233+134 & QSO   & 45.33 & $8.04_{-0.10}^{+0.08}$ & 6 & 0.67$\pm$0.23 &  0.28$\pm$0.28 & 1 \\
 3C 390.3    & Sy1   & 43.99 & $8.85_{-0.11}^{+0.09}$ & 6 & 1.80$\pm$0.22 & -0.61$\pm$0.15 & 1 \\
 I Zw 1      & Sy1   & 44.80 & $7.44_{-0.12}^{+0.09}$ & 6 & 0.85$\pm$0.13 & -0.85$\pm$0.28 & 2 \\
 Mrk 509     & Sy1   & 44.28 & $8.16_{-0.04}^{+0.04}$ & 7 & 0.84$\pm$0.14 &  0.66$\pm$0.35 & 2 \\
 Mrk 573     & Sy1   & 44.40 & $7.28_{-0.10}^{+0.08}$ & 8 & 0.98$\pm$0.24 & -2.35$\pm$0.14 & 3 \\
 Mrk 841     & Sy1.5 & 44.29 & $8.52_{-0.10}^{+0.08}$ & 6 & 1.07$\pm$0.25 &  0.05$\pm$0.35 & 2 \\
 NGC 3227    & Sy1.5 & 42.38 & $7.63_{-1.9}^{+1.1}$ & 7 & 0.98$\pm$0.24 & -2.55$\pm$0.21 & 4,9 \\
 NGC 3783    & Sy1   & 43.26 & $7.47_{-0.09}^{+0.07}$ & 7 & 0.51$\pm$0.14 & -0.34$\pm$0.35 & 2 \\
 NGC 4593    & Sy1   & 43.09 & $6.73_{-0.09}^{+0.03}$ & 7 & 0.34$\pm$0.13 & -3.44$\pm$0.45 & 2,9 \\
 NGC 5548    & Sy1   & 43.51 & $7.83_{-0.02}^{+0.02}$ & 7 & 0.73$\pm$0.10 & -0.81$\pm$0.26 & 5,9 \\
 NGC 7469    & Sy1   & 43.72 & $7.09_{-0.05}^{+0.05}$ & 7 & 0.26$\pm$0.06 & -1.16$\pm$0.43 & 2 \\
 \hline
 \multicolumn{8}{l}{(1) \rule{0pt}{11pt}\footnotesize This paper;
                    (2) \rule{0pt}{11pt}\footnotesize Smith et al. (2002);
                    (3) \rule{0pt}{11pt}\footnotesize Nagao et al. (2004);
                    (4) \rule{0pt}{11pt}\footnotesize Axon et al. (2008)}\\
 \multicolumn{8}{l}{(5) \rule{0pt}{11pt}\footnotesize Goodrich and Miller (1994);
                    (6) \rule{0pt}{11pt}\footnotesize Vestergaard and Peterson (2006);
                    (7) \rule{0pt}{11pt}\footnotesize Peterson et al. (2004);}\\
 \multicolumn{8}{l}{(8) \rule{0pt}{11pt}\footnotesize Satyapal et al. (2005);
                    (9) \rule{0pt}{11pt}\footnotesize Wu and Han (2001).}\\
 \end{tabular}
\end{table}

\begin{figure}
 \centerline{\includegraphics[height=21cm]{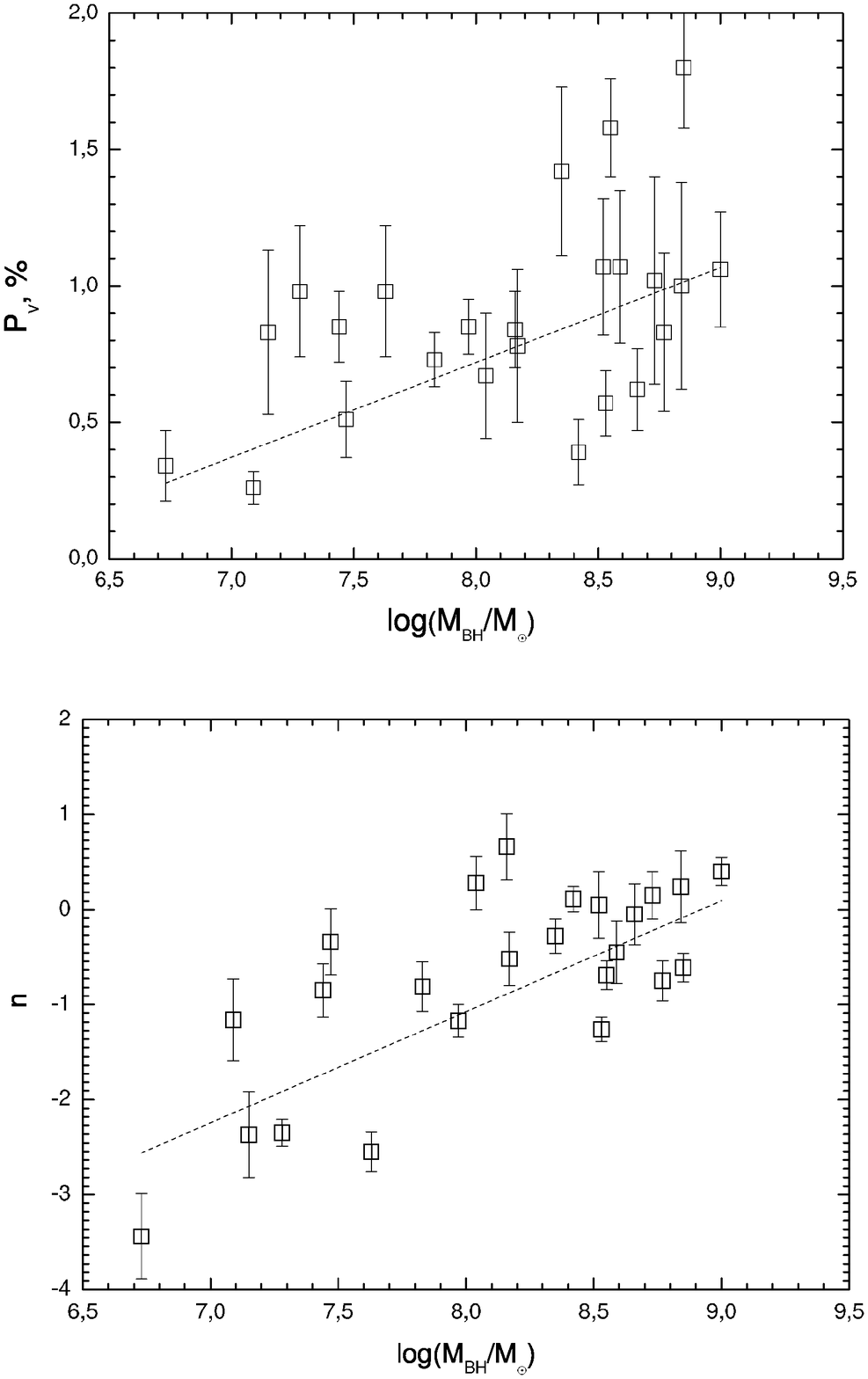}}
 \caption{Linear polarization and power-law index n versus black hole masses from the data of Table 2 (see also the text)}
\end{figure}

\section{The main technique of determining the magnetic field strength and distribution in an accretion disk}

The technique of determining the magnetic field strength and topology from spectropolarimetric observations is based on the works by Gnedin and Silant'ev (1984, 1997), Silant'ev et al. (2009), and Dolginov et al. (1995). Anisotropy in the hot-gas density distribution in the region of scattering of electromagnetic radiation is known to produce polarized emission due to the Thomson scattering of photons by hot-gas electrons. At the same time, the presence of a magnetic field in the scattering medium leads to anisotropy in the propagation and absorption of electromagnetic radiation. If, however, the magnetic field is not strong enough ($B < 10^6$ G ) to provide optical anisotropy of the medium, then the real optical anisotropy can result from the Faraday rotation of the polarization plane on the photon mean free path in the medium (Gnedin and Silant'ev 1984). The formula for the Faraday rotation angle on the photon mean free path can be represented as (Dolginov et al. 1995; Gnedin and Silant'ev 1997)

\begin{equation}
 \Psi({\bf n},{\bf B}) = \frac{1}{2} \delta \tau \cos{\theta};\,\,
 \delta = \frac{3\lambda}{4\pi r_e} \frac{\omega_B}{\omega} \approx 0.8 \lambda^2[\mu m] B[G]
 \label{eq1}
\end{equation}

\noindent where ${\bf n}$ specifies the direction of the electromagnetic radiation, ¨ is the angle between the directions of ${\bf n}$ and the magnetic field ${\bf B}$, $\tau = \sigma_T N_e l$ is the Thomson optical depth ($N_e$ is the number density of free electrons, $l$ is the geometrical length of the electron scattering region, $\sigma_T = (8\pi / 3)(e^2 / m_e c^2)^2$ is the Thomson scattering cross section), $\omega_B = e B / m_e c$ is the cyclotron frequency, $\omega$ is the radiation frequency, and $\lambda = 2\pi c /\omega$ is the radiation wavelength in $\mu$m.

The Faraday rotation of the polarization plane on the photon mean free path leads to a strong wavelength dependence of the polarization and the position angle of the polarized radiation emerging from a plane-parallel, optically thick plasma atmosphere (Silant'ev 2002; Silant'ev et al. 2009):

\begin{equation}
 P_l({\bf B},{\bf n}) = \frac{P_l(0,\mu)}{\sqrt{1 +
 \delta^2\cos{\theta}^2}};\,\, \tan{2\chi} =
 \frac{U_{\lambda}}{Q_{\lambda}} = \delta \cos{\theta}
 \label{eq2}
\end{equation}

\noindent Here, $B\cos{\theta} = {\bf B n}$ and $\mu = \cos i$, where $i$ is the inclination of the accretion disk, $U$ and $Q$ are the Stokes parameters. Using Eq.(\ref{eq2}) to analyze and interpret spectropolarimetric observations requires knowing the actual magnetic field topology in the accretion disk.

The axial symmetry of the accretion disk allows one to separate the effects attributable to the horizontal (within the accretion disk) and vertical (poloidal) components of the global magnetic field. Averaging Eq.(\ref{eq2}) over the azimuth angle, we obtain (Silant'ev et al. 2009) the following formulas for the Stokes parameters:

\[
 \langle Q\rangle = Q(0,\mu) \frac{2}{\pi} \int_0^{\pi/2} d\varphi
 \frac{1+a^2+b^2\cos{\varphi}^2}{(1+a^2+b^2\cos{\varphi}^2) -
 (2ab\cos{\varphi})^2}
\]

\begin{equation}
 \langle U\rangle = Q(0,\mu) \frac{2a}{\pi} \int_0^{\pi/2} d\varphi
 \frac{1+a^2-b^2\cos{\varphi}^2}{(1+a^2+b^2\cos{\varphi}^2) -
 (2ab\cos{\varphi})^2}
 \label{eq3}
\end{equation}

\noindent where the depolarization parameters $a$ and $b$ are

\begin{equation}
 a = 0.8 \lambda^2 B_z \mu;\,\,
 b = 0.8 \lambda^2 B_{\bot} \sqrt{1-\mu^2}
 \label{eq4}
\end{equation}

\noindent Here, $B_z$ is the component of the global magnetic field of the accretion disk along the normal to the disk surface and $B_{\bot}$ is the magnetic field inside the disk. Recall that $\mu = \cos{i}$ and $i$ is the angle between the line of sight and the normal.

In what follows, we will assume a power-law dependence of the magnetic field strength on the accretion disk radius with an exponent $s$:

\begin{equation}
 B(R) = B_H (R_H / R)^{s}
 \label{eq5}
\end{equation}

\noindent where $B_H$ is the magnetic field strength near the supermassive black hole horizon radius $R_H$. The exponent in the power-law dependence of the magnetic field on radius s can take on any values. However, for the model of an advection-dominated accretion flow (ADAF) and a standard disk with equal magnetic and kinetic pressures, its value is $s = 5/4$ (Pariev et al. 2003).

Significantly, this value can be determined directly from spectropolarimetric observations provided that the radial dependence of the accretion disk temperature, i.e., $T_e(R)\sim R^{-p}$, is known. For the standard model (Shakura and Sunyaev 1973), $p = 3/4$. According to Silant'ev et al. (2009), in the case of strong Faraday depolarization, Eqs.(\ref{eq3}) lead to the following wavelength dependence of the polarization:

\begin{equation}
 P_l \sim \frac{P_l(0,\mu)}{B_{z,\bot} \lambda^2} \sim
 \lambda^{(s /p - 2)}
 \label{eq6}
\end{equation}

Recall that the spectral distribution of the radiation
from a standard accretion disk is

\begin{equation}
 F_{\nu} \sim \nu^{3 - 2/p}
 \label{eq7}
\end{equation}

In the case of a standard accretion disk ($p = 3/4$, $s = 5/4$), we obtain $P_l \sim \lambda^{-1/3}$, $F_{\nu} \sim \nu^{1/3}$, and $P_{\nu} F_{\nu} \sim \nu^{2/3}$.
When the radial temperature distribution is $T_e \sim R^{-1/2}$, we have $P_l \sim \lambda^{1/2}$, $F_{\nu}
\sim \nu^{-1}$, and $P_{\nu} F_{\nu} \sim \nu^{-3/2}$. This case corresponds to a situation when the characteristic radius of the accretion disk from where radiation with wavelength $\lambda$ corresponding to an effective black body temperature $T_e (R)$ emerges is $R(\lambda) \sim \lambda^2$ instead of the relation $R(\lambda) \sim \lambda^{4/3}$ characteristic of the standard Shakura-Sunyaev model. This means that the spectral distribution of the accretion disk radiation corresponds to a radial dependence of the temperature $\sim R^{-0.5}$ irrespective of the accretion disk thickness.

Generally, calculating the polarization and position angle using Eqs.(\ref{eq3}) requires numerical simulations (Silant'ev et al. 2009). However, simple analytical formulas useful for comparison with spectropolarimetric observations can be derived in several cases. Consider two important cases of the global magnetic field topology in an accretion disk.

Thus, if the vertical magnetic field component dominates, i.e., $B_z \gg B_{\bot}$, then the polarization and position angle are

\begin{equation}
 P_l({\bf B},\mu) = \frac{P_l(0,\mu)}{\sqrt{1+a^2}},\,\,
 \tan 2 \chi = a
 \label{eq8}
\end{equation}

If the horizontal magnetic field component dominates, i.e., when $B_{\bot} \gg B_z$), we have:

\begin{equation}
 P_l({\bf B},\mu) = \frac{P_l(0,\mu)}{\sqrt{1+b^2}},\,\,
 \chi = 0
 \label{eq9}
\end{equation}

Strong turbulence can emerge in a magnetized accretion disk. In this case, the asymptotic expressions for the polarization and position angle are

\begin{equation}
 P_l({\bf B},{\bf n}) = \frac{P_l(0,\mu)}{\sqrt{(1+c)^2 + \delta^2 \cos{\theta}^2}},\,\,\,
 \tan{2\chi} = \frac{U}{Q} = \frac{\delta \cos{\theta}}{1+C}
 \label{eq9_1}
\end{equation}

The parameter $C$ emerges in a turbulent medium (Silant'ev 2005) and characterizes a new effect - an additional decrease in Stokes parameters $Q$ and $U$ due to incoherent Faraday rotation on small-scale turbulent vortices. The expression for the parameter $C$ is (Silant'ev 2005)

\begin{equation}
 C = 0.21 \tau_1 \lambda^4 \langle B'^2 \rangle
 \label{eq9_2}
\end{equation}

\noindent Here, $\tau_1$ is the characteristic optical depth of a turbulent vortex and $B'$ is the amplitude of the turbulent magnetic field. Interestingly, in this case, the depolarization parameter $\sim \lambda^4$. This leads to an asymptotic wavelength dependence of the polarization $P_l \sim \lambda^{-2}$ if the amplitude of the turbulent magnetic field does not depend on distance in the accretion disk.

The value of $\chi = 0$ corresponds to a situation where the polarization plane coincides with the accretion disk plane. Formulas (\ref{eq8}) and (\ref{eq9}) are convenient for comparison with observational data.

Recall that the quantity $P_l(0,\mu)$ is well known from the theory of polarized radiation transfer in a plane-parallel atmosphere (Sobolev 1949; Chandrasekhar 1950). For a plane-parallel atmosphere seen edge-on ($i = 90^{\circ}$), the maximum polarization in the case of multiple electron scattering is 11.7\%.

The above formulas now allow us to determine the magnetic field strength and topology in the radiation generation region for the objects observed in our program.

\section{Main results of our observations: determination of physical parameters for accretion disks around supermassive black holes}

Let us analyze the spectropolarimetric observations of specific objects in our program.

The results of our analysis are presented in Tables 3 and 4. Their respective columns give the object names, power-law exponents $p$ in the radial dependence of the temperature in the accretion disk, power-law exponents $s$ in the radial dependence of the magnetic field strength in the accretion disk, and magnetic field strengths $B(R_{\lambda})$ in the region of the observed radiation wavelength.

In principle, the magnetic field strength in AGNs can be determined from the following data: (1) the accretion disk inclination to the line of sight; (2) the polarization and position angle of the radiation emerging from an optically thick atmosphere due to electron scattering (the Sobolev.Chandrasekhar theory); (3, 4) the observed wavelength dependences of the polarization and position angle; and (5) the theoretical estimates of the radiative efficiency $\varepsilon$ of the accretion process obtained by Shapiro (2007) and other authors (see the monograph by Novikov and Thorne 1973), which depends on the black hole rotation rate.

A detailed technique for calculating the AGN magnetic field is presented in Silant'ev et al. (2009).

As regards the magnetic field strength in a marginally stable orbit and near the event horizon, it can be estimated within the framework of the magnetic coupling model (Ma et al. 2007). This quantity depends on the ratio of the parameters characterizing the ratio of the magnetic and kinetic energy densities near the event horizon and the radiative efficiency of the accretion process dependent on the black hole rotation rate, i.e., the ratio $k/\varepsilon$. Therefore, the choice of an accretion disk model cannot be completely avoided. For $\varepsilon = 0.32$ corresponding to $a_{\ast} = 0.998$ (Novikov and Thorne 1973; Shapiro 2007), the quantity $k$ takes on its maximum value closest to the equality of the magnetic and kinetic energy densities. It is this value that best corresponds to the magnetic coupling model.

It is interesting to note that the set of observational data, in principle, allows the parameters $k$ and $\varepsilon$  to be determined independently, but under the condition that we can determine the position angle precisely relative to the accretion disk itself. Since, unfortunately, this cannot yet be done, here we will not estimate the magnetic field in the immediate vicinity of the black hole.

\begin{table}
 \caption[]{Physical parameters of the accretion disk obtained from our spectropolarimetric observations at the 6-m BTA telescope (SCORPIO) and published spectroscopic data.}
 \centering
 \begin{tabular}{l c c c}
 \hline
 Object & $p$ & $s$ & $B(R_{\lambda})$[G] \\
 \hline
 PG 0007+106 & 1/2 & 1   & 2.43 \\
 PG 0026+129 & 3/4 & 5/4 & 1    \\
 PG 0049+171 & 3/4 & 5/4 & 13   \\
 PG 0157+001 & 3/4 & 5/4 & 98   \\
 PG 0804+761 & 3/4 & 3/2 & 3.4  \\
 PG 0844+349 & 3/4 & 1   & 37   \\
 PG 0953+414 & 3/4 & 1   & 300  \\
 PG 1116+215 & 3/4 & 3/4 & 100  \\
 PG 2112+059 & 3/4 & 2   & 14.4 \\
 PG 2130+099 & 1/2 & 1   & 27   \\
 PG 2209+184 & 1/2 & 3/4 & 16   \\
 PG 2214+139 & 1/2 & 5/4 & 2.8  \\
 PG 2233+134 & 3/4 & 3/2 & 0.37 \\
 3C 390.3    & 3/4 & 1   & 6.4  \\
 \hline
 \end{tabular}
\end{table}

\begin{table}
 \caption[]{Magnetic field strengths and accretion disk parameters for some central black holes under the same conditions as those in Table 3}
 \centering
 \begin{tabular}{l c c c}
 \hline
 Object & $p$ & $s$ & $B(R_{\lambda})$[G] \\
 \hline
 I Zw 1   & 1/2 & 1   & 11   \\
 Mrk 509  & 1/2 & 3/4 & 4.8  \\
 Mrk 841  & 1/2 & 1   & 3.0  \\
 NGC 3227 & 3/4 & 1   & 15.7 \\
 NGC 3783 & 3/4 & 1.5 & 41   \\
 NGC 5548 & 3/4 & 1   & 7.6  \\
 NGC 7469 & 3/4 & 1   & 100  \\
 \hline
 \end{tabular}
\end{table}

To determine the magnetic field strength in the region where the radiation is generated from the degree of its Faraday depolarization, we must know the accretion disk inclination i with respect to the line of sight. Unfortunately, there are no reliable methods for determining the inclination i at present. For most of the AGNs listed in Tables 1 and 2, we used the available published data (Braatz et al. 1997; Crummy et al. 2005; Ho et al. 2008). For instance, according to Crummy et al. (2005), the inclinations for the objects PG~0157+001 and PG~0844+349 observed at the 6-m BTA telescope are $i = 73^{\circ}$ and $i = 59^{\circ}$, respectively, with an error $\leq 10^{\circ}$. In some cases, we used $i = 60^{\circ}$. In this case, without allowance for the depolarization, the polarization should be $P_{\nu} \approx 2\%$.

Table 3 does not include the observed object PG~1022+519 for which a strong wavelength dependence of the polarization was obtained, $P_l \sim \lambda^{-2.37}$. Such a strong dependence is obtained in the model of a turbulent accretion disk (see Eqs. (\ref{eq9_1}) and (\ref{eq9_2})). In accordance with (\ref{eq9_1}) and (\ref{eq9_2}), we get the following estimate of the magnetic field turbulence amplitude in the radiation region: $B' \approx 3.5$G.

For the objects included in Table 3, we also assumed that the direction of the global magnetic field makes an angle of $60^{\circ}$ with the accretion disk surface. Many numerical simulations (Lyutikov 2009; Sadowski and Sikora 2010; Cao 1997) show that precisely this direction of the global magnetic field is realized at large distances from the horizon for the Kerr metric.

Let us briefly discuss the dependences presented in Figs.1 and 2. The weak dependence of the polarization on the central supermassive black hole mass is attributable to the same weak dependence of the depolarization parameters for $s = 1,\,5/4$. For instance, for $s = 1$,

\begin{equation}
 a \sim b \sim \left(\frac{M_{\odot}}{M_{BH}}\right)^{1/3}
 L_{bol}^{1/6} \lambda^{1/3}
 \label{eq9_3}
\end{equation}

A slightly different situation takes place for the power-law exponent $n$ in the wavelength dependence of the polarization. Since the wavelength dependence of the polarization arises only in the case of noticeable Faraday depolarization, this requires that the depolarization parameters $a$ and $b$ be at least of the order of unity, i.e., $a \geq 1$ and $b \geq 1$. The wavelength entering into Eq.(\ref{eq4}) and expressed in $\mu$m is always less than unity ($\lambda < 1$) in the observed wavelength range. Hence, a situation arises where for negative values, $n < 0$, the condition $a > 1$ is also achieved at lower values of $M_{BH} / M_{\odot}$. In contrast, the case of $n > 0$ requires higher values of $\lambda$, which can also be realized at higher values of $M_{BH} / M_{\odot}$.

For $n > 0$, the degree of polarization grows with wavelength, approaching the Chandrasekhar limit, which is reached faster precisely at high values of $M_{BH} / M_{\odot}$. This result is obtained if the magnetic field strength $B_H$ decreases with increasing ratio $M_{BH} / M_{\odot}$. Curiously, it is this dependence of the magnetic field strength on the black hole mass that takes place in the magnetic coupling model.

Another source of polarized radiation in AGNs is a hot scattering corona that can exist above the accretion disk surface. In our published paper (Piotrovich et al. 2010), we calculated the degree of polarization and its wavelength dependence in the case where the polarization results from the scattering of accretion disk radiation in a hot corona produced by a plasma outflow from the accretion disk (disk wind). The calculations were performed for the case where the magnetic field is described by the Parker model. The strong dependence of the polarization, which is not described by a power law, is determined mainly by the toroidal magnetic field component. Another important circumstance that allows us to exclude the situation with a hot corona from consideration is that a fairly high plasma column density in the hot corona, $N_H \geq 6\times 10^{23} cm^{-2}$, is required for the emergence of a noticeable polarization. Such a high column density is not characteristic of the observed AGNs but corresponds to the case of a hot corona optically thick with respect to Compton scattering.

\section{Conclusions}

Using the SCORPIO universal focal reducer at the 6-m BTA telescope, we performed spectropolarimetric observations of active galactic nuclei and quasars. The derived wavelength dependences of the polarization and position angle were compared with the results of theoretical calculations based on the theory of multiple scattering with allowance made for the rotation of the polarization plane in the process of electron scattering (Dolginov et al. 1995; Gnedin and Silant'ev 1997; Gnedin et al. 2006; Silant'ev et al. 2009).

As a result, we determined the magnetic field strengths and radial distributions in an accretion disk around a supermassive black hole within the framework of traditional accretion disk models. For a number of active galactic nuclei (blazars), the hot plasma cloud (corona) surrounding the disk plays a more important role than the accretion disk. This special case was considered by Piotrovich et al. (2010).

\section*{Acknowledgements}

This work was supported by Program ü 4 of the Presidium of the Russian Academy of Sciences, the Program of the Division of Physical Sciences of the Russian Academy of Sciences, the ''Scientific and Scientific-Pedagogical Personnel of Innovational Russia'' Federal Goal-Oriented Program, and the ''Leading Scientific Schools'' Presidential Program for 2010-2011 (NSh-3645.2010.2).

\smallskip
\raggedleft
{\it Translated by G. Rudnitskii.}

\end{document}